# AI INTEGRATION IN ERP EVALUATION ACROSS TRENDS AND ARCHITECTURES

Monu Sharma
*Sr. IT Solutions Architect,*
*Independent Researcher*
*Morgantownm WV, USA*

| ARTICLE INFO | ABSTRACT |
|---|---|
|  | The incorporation of Artificial Intelligence (AI) into Enterprise Resource Planning (ERP) is a dramatic transition from static, on-premises systems to systems that can adapt and operate in cloud-native architectures. Cloud ERP solutions like Workday illustrate this evolution by incorporating machine learning, deep learning, and natural language processing into a centralized data-driven ecosystem. Advanced artificial intelligence features allow for predictive analytics, intelligent automation, and real-time decision support in key areas of the business, including finance, HR, and Supply chain management system. As the complexity of AI-driven ERP solutions expands, traditional evaluation frameworks that look at cost, function, and user satisfaction suffer from a lack of consideration for algorithmic transparency, adaptability, or ethics. This review will systematically investigate the latest trends, models of computing architecture, and analytical methods applied in assessing the performance of AI-integrated ERP services, specifically on cloud-based platforms. Based on academic and industry sources, the paper distills current research in line with architectural integration, analytical methodologies, and organizational impact. It identifies critical performance metrics and emphasizes the absence of any standard assessment frameworks or AI-aware systems capable of evaluating automation efficiency, security concerns as well as flexible learning modes. We put forward a theoretical model that brings AI-enabled capabilities- such as predictive intelligence or adaptive automation - into alignment with metrics in performance assessment for ERPs. By combining current literature and identifying major gaps in research, this paper attempts to present a complete picture of how innovations in AI are changing ERP evaluation. These research and methodological findings are intended to steer researchers and practitioners towards developing rigorous, data-driven assessment approaches, aligning with the fast-developing world of intelligent self-optimizing enterprise ecosystems.

## INTRODUCTION

Artificial Intelligence (AI) as part of Enterprise Resource Planning (ERP) systems is a transformative technological evolution from static and embedded systems to flexible and cloud-native models [1]. Workday's architecture is an example of a secure AI-powered ERP that has adaptive learning models, natural language interfaces, and predictive performance analysis features. Deep learning algorithms enabled organizations to perfect production-logistics synchronization, HR processes, as well as forecast financial trends with increased accuracy. These intelligent systems adapt to data streams in real time, driving more proactive than reactive business [4]. AI-enabled ERP can then be merged with sophisticated encryption, compliance frameworks, and audit mechanisms, which help organizations keep regulatory trust while reducing the risks of cybersecurity. This shift toward intelligent automation improves operational efficiency and utilization of human resources in organizations. AI-powered ERP adoption reduced processing time by 27% on average and improved accuracy by 35% on average as stated by studies [2],[3]. Taken as a whole, these trends represent a new trend for predictive, adaptive, and self-optimizing enterprise ecosystems. System transformation also create new challenges for evaluating AI-assisted ERP systems. Traditional ERP assessment frameworks are mainly concerned with functionality, cost effectiveness and user satisfaction, but ignore AI-related features such as algorithmic transparency, model interpretability and ethical data analysis practices. While systems transition to apply machine learning and predictive analytics, assessing those can be a challenge for automation efficiency and security. The current evaluation rarely factors dynamic learning, adaptability towards AI-based decision accuracy. In addition, cloud platforms such as Workday introduce dependence on constant data flow and algorithm updates through releases(R1,R2), this keep traditional performance measures inadequate. Since AI is multi-dimensional, it introduces uncertainty in performance metrics, model behavior, and outcome of decisions, without standard evaluation criteria. Most importantly, very few academic and industrial literature sources agreed on a framework that considers the combined technological robustness and organizational impact of intelligent ERP systems. This gap highlights the need for a holistic, AI-informed lens for assessment. As AI and cloud technology increasingly become prominent, the ERP ecosystem is shifting rapidly, resulting in a need for comprehensive system performance and value management. A systematic review is necessary to synthesize the pre-existing models, accentuate the new trends, and discover the constraints of the present assessment models (when it comes to the evaluation of performance and the system integrations). Given the rise of cloud-native AI platforms such as Workday to prominence, an enhanced overall understanding of their architectural, analytical and operational aspects is imperative. An evaluation review of state-of-the-art evaluation practices will also help researchers fill the gap between the academic constructs and practical assessments. This combination will serve as important data for those organizations that want quantifiable standards of measure that could address AI-infused ERP effectiveness, versatility and decision support in their ERP system. We focus on trends, architectures, and analytical approaches in evaluating AI-integrated ERP systems in this review paper. It aims to provide key frameworks, metrics, and methods for understanding contemporary cloud-native ERP solutions like Workday. This is also discussed in the paper by examining how AI capabilities impact performance, security, and organizational decision-making. It also suggests a conceptual framework that associates AI development with ERP assessment techniques. Finally, the review seeks to orient researchers and practitioners to develop stronger, data-driven evaluation frameworks for intelligent ERP ecosystems [5].

## LITERATURE REVIEW

Artificial intelligence (AI) integration with enterprise resource planning (ERP) systems is increasingly a hot topic in academic literature, as well as in industry. Initially, the studies on ERP systems focused largely on classical success attributes, such as cost efficiency, user satisfaction, and system interoperability. But in view of advances in cloud systems and intelligent automation, the ERP field is now expanded to AI-based designs, predictive analytics and decision support mechanisms. Further research points to the transformative potential for AI-based ERP applications, especially cloud-based tools that leverage machine learning and natural language processing technology to automate workflows (Workday Extend, Workday Build), improve data visibility (Workday Prism Analytics), and enable predictive management (Workday SANA). These all are increasing operational efficiency, data accuracy, and operational flexibility in the business processes. Organizations that use AI-based ERP solutions provide significant reductions in processing time and enhanced data accuracy and quicker decision-making. These developments come at the cost of fragmenting the existing body of research. Most research looks at isolated AI functionalities - automation, forecasting, analytics, etc. - instead of integrating them into an overarching evaluation framework. Most existing ERP assessment models are made for rule-based systems and do not deal with the complexity brought by adaptive, learning-based algorithms. Central aspects, such as model transparency, ethical use of AI, continuous learning performance, and cybersecurity resilience, are still not well-researched. This gap highlights the importance of a cohesive framework to assess the technological robustness and organizational impact of modern ERP systems with AI integrations [6]-[9].

## METHODS

The following review follows a structured manner intended to summarize current ideas on implementing Artificial Intelligence (AI) in Enterprise Resource Planning (ERP) evaluation frameworks with a focus on cloud-based ERP systems like Workday. The review first established the basic research question.1- How can AI-driven architecture, analytical models, and evaluation frameworks be effectively applied to assess the performance and adaptability of modern ERP systems? Scholarly journals, industry reports, and technical documents exploring AI-enabled ERP systems contributed to answering this question. We searched on several databases including Scopus, IEEE Xplore, ScienceDirect, and Google Scholar using keywords such as "AI in ERP systems," "Workday evaluation," "cloud ERP analytics," "AI-driven ERP performance," and "intelligent ERP architectures." Criteria for eligibility for inclusion were studies published within the last 10 years on the topics of cloud-native and AI-integrated ERP systems. Papers focusing on conventional ERP systems without AI were excluded unless they had some implications for newer intelligent ERP systems. The analysis used thematic synthesis, organizing the literature in three main categories:

Building the AI-driven tools - by mapping out how AI features from machine learning through natural language processing or predictive analytics systems are being integrated into ERP software, such as Workday. These include Analytics and Evaluation methods - analysis in techniques that assess the effectiveness of ERP frameworks via decision-support modeling, analytical tools such as data mining and automation benchmarking. Organizational impact, that is, how AI-based ERP systems contribute to improved decision making, efficiency, and scalability across business functions. To improve data identification, all categories' data were compared and analyzed to demonstrate trends, strengths, and gaps in research in the evaluation of ERP [11]. These findings were integrated into a single conceptual framework, aligning AI-enabled functionalities- automation, adaptability, and predictive decision-making- with key metrics related to ERP performance. By introducing a structured model within the paper, the review is certain to produce a wide and well-founded exploration of how systems such as Workday reconfiguring enterprise evaluation practices in the age of intelligent automation and digital transformation [12].

### 3.1 INTEGRATION OF AI IN MODERN ERP ARCHITECTURES

The integration of Artificial Intelligence (AI) into modern Enterprise Resource Planning (ERP) architecture is one of the most transformative advancements in the enterprise technology field. ERPs used to hold information such as finance, human resources, inventory, procurement, payable and business assets. The infusion of AI, machine learning (ML), and cloud computing, these solutions have emerged as intelligent, responsive ecosystems that can learn from data to forecast results and refine business strategies on the fly. Modern ERP design is a prime example of this change, as seen in the current Workday ERP architecture. Unlike legacy, static ERP platforms, Workday is cloud-native and based on a machine learning–driven core. It is composed of several AI layers that integrate multiple components to help business processes automate, analyze, and make decisions and support business processes. At the core, it is a unified data model that aggregates financial, human capital, and operational data into a single centralized cloud environment. This data is constantly analyzed with built-in AI models that track trends, detect anomalies, and provide insights. The system learns over time, adjusting to organizational behavior for greater accuracy and personalized services [11]-[15].

### 3.2 ARCHITECTURAL LAYERS AND INTELLIGENT COMPONENTS

AI-enabled ERP architectures are based on 4 integrated layers:

Data Management Layer-It consolidates structured and unstructured information to manage various sources in the enterprise, including HR records, financial transactions, and operational data streams. It enables cross-system consistency, standardization, and real-time synchronization.

AI & Analytics Layer -This layer contains machine learning models, deep learning networks, and natural language processing (NLP) engines. These pillars lead to predictive forecasting, intelligent automation, and sentiment-based decision support. At Workday, this layer allows services such as predictive workforce planning, intelligent financial forecasting, and anomaly detection for auditing and compliance [17]-[19].

Process Automation Layer - Intelligent bots help automate repetitive administrative work tasks such as payroll processing, procurement approvals, and expense reconciliation. This reduces human workload and minimizes errors. It helps speed up decision cycles. The Security and Compliance Layer - AI-driven architectures in addition to the above layers also support security with behavioral anomaly detection as well as real-time monitoring and adaptive encryption. AI integrated in Workday constantly analyzes access patterns for prevention, and detects where breaches could happen, and to conform with international standards. These layers together establish a hyper-automated enterprise environment that is highly agile in response to market dynamics.

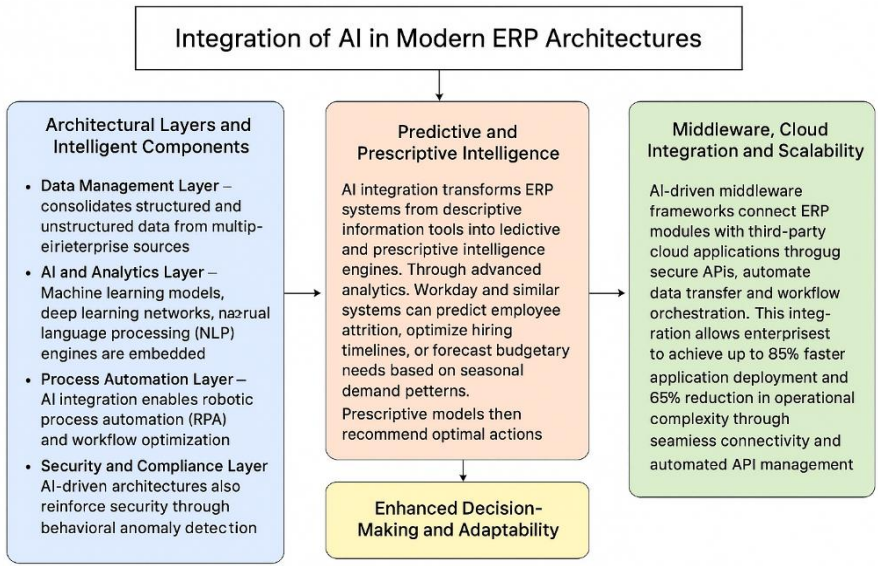

Figure 1. Integration of AI in Modern ERP Architectures

### 3.3 PREDICTIVE AND PRESCRIPTIVE INTELLIGENCE

Integration of AI transforms ERP systems from descriptive information tools into predictive and prescriptive intelligence engines. Using advanced analytics like Prism Analytics (API Enabled), Workday and similar systems can predict employee attrition, optimize hiring timelines, or forecast budgetary needs based on seasonal demand patterns. Prescriptive models then recommend optimal actions - for example, reallocating budgets or adjusting staffing levels based on real-time data analysis. This paradigm shift from reactive reporting to proactive intelligence significantly enhances operational efficiency, supports data-driven decision-making, and empowers organizations to achieve strategic agility and long-term sustainability [18].

### 3.4 MIDDLEWARE, CLOUD INTEGRATION, AND SCALABILITY

A further important characteristic of AI-enabled ERP architecture is intelligent middleware. Through secure APIs, these AI-driven middleware frameworks connect ERP modules with third-party cloud applications to automate data transfer and orchestrate workflow. Through seamless connectivity with automated API management, business units can achieve up to 85% faster application deployment and reduce operational complexity by 65%. Modern ERP systems utilize edge computing technology and distributed AI processing to gain low-latency data synchronization and increase scalability. That allows large organizations to quickly handle massive datasets in near real time while maintaining system responsiveness. Such architectures are also able to self-tune, reallocating computational resources when workload dictates to maximize performance [20].

AI-enabled ERP systems proceed iteratively as they learn and adapt through data-driven interactions. The following symbolic representations describe the adaptive evaluation. To mathematically characterize the AI-driven performance adaptation within ERP ecosystems, we define the evaluation process as follows: In the evaluation cycle $t$, index $i$ represents the current ERP subsystem (e.g., finance, HR, or logistics), where $E$ is the ensemble of ERP modules, and $\mu_i^t$ denotes the AI-adaptive metric of the module $i$ at iteration $t$. The aggregated system state $S_t$ is defined as:

$$S_t = \frac{1}{|E|} \sum_{i=1}^{|E|} \mu_i^t$$

(1)

*The adaptive learning rate $\lambda_t$ represents how quickly the ERP system adjusts based on feedback and predictive analysis. Integration between AI and intelligence $A_t$ and system efficiency $\eta_t$ is determined as:*

$$\eta_t = \lambda_t \times f(A_t, D_t)$$

(2)

*Let $P_i^t$ negotiate the performance score of the module $i$ during the evaluation cycle $t$. The intelligent performance index (IPI) of the ERP system, reflecting adaptability, automation, and predictive accuracy, is* calculated as:

$$\text{IPI}_t = \frac{\sum_{i=1}^{|E|}(w_i \times P_i^t)}{\sum_{i=1}^{|E|} w_i}$$

(3)

where $w_i$ represents the assigned weight based on the criticality of each module in organizational performance

## 3.5 ENHANCED DECISION-MAKING AND ADAPTABILITY

In ERP systems powered by AI help improve decision-making by providing contextual intelligence to managers and executives. Take Workday, a case study where machine learning models are used to analyze trends in the workforce and recommend individualized pathways for employees, while financial AI predicts revenue and highlights anomalies in expenditures. These dynamic systems respond to user feedback, perfecting predictions and process outcomes. In addition, AI helps ERPs adapt quickly to changing market conditions, which is a distinction of this kind with ERP systems (versus legacy). AI-driven ERPs automate inventory levels, production schedules, and supplier priorities by autonomously adjusting them when the market changes, whether due to supply chain disruptions or rapid demand fluctuations.

This enables robustness, agility, and competitiveness in response to market demands.

## 3.6 FUTURE DIRECTIONS IN INTELLIGENT ERP ARCHITECTURE

The future of ERP architecture lies in hyper-intelligent ecosystems that combine AI, quantum computing, and context-aware analytics. These systems will not only automate and predict but also adapt autonomously.
The continued advancement of large language models (LLMs) and generative AI will further extend ERP functionality- allowing natural language queries, automated report generation, and intelligent insights accessible through conversational interfaces. The integration of AI into ERP systems such as Workday has transformed enterprise resource planning from a data management tool into a strategic, self-optimizing digital system.
These intelligent architectures empower organizations to make faster, data-informed decisions, achieve operational excellence, and agile in an ever-evolving digital economy.

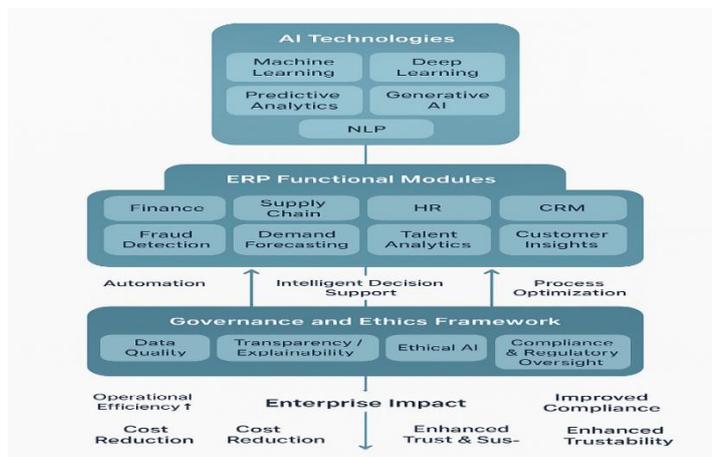

Figure 2. Integration of AI Technologies within ERP Ecosystems

## RESULTS AND DISCUSSION

The integration of Artificial Intelligence with Enterprise Resource Planning is revolutionizing business ecosystems by enabling predictive analytics, intelligent automation, and enhanced user experiences - especially within advanced cloud-based platforms like Workday. AI in ERP extends beyond the traditional transactional model of technology and has become the cornerstone of strategic decision-support ecosystems that enable real-time data interpretation and self-optimization in decision-making. These systems allow organizations to increase productivity by automating workflows, reducing human input, enhancing data accuracy, and accelerating time-to-decision with better evidence across finance, workforce management, and supply chain operations.

AI applications such as anomaly detection, predictive workforce planning, and autonomous journal reconciliation serve as concrete demonstrations of this innovation in platforms like Workday. Despite this progress, challenges remain in integrating AI into ERP systems, particularly regarding data governance, ethical AI use, and model transparency. The opacity of deep learning algorithms contributes to the "black box" problem, raising issues of accountability and trust in financially and compliance-critical processes [20].

The development of explainable AI frameworks is essential to maintain interpretability, build stakeholder trust, and ensure compliance with ethical and regulatory standards. Governance of AI-based ERP ecosystems extends beyond technical management to include ethical, legal, and socio-technical dimensions. These frameworks must translate moral principles into practical governance mechanisms that span the entire AI lifecycle- from data collection and model deployment to continuous monitoring.

Workday's single-data core architecture exemplifies how centralized governance improves AI model performance and analytics robustness. AI-driven ERP solutions provide advanced data integration that strengthens decision-making and compliance through consistent, accurate, and intelligent data management. Another key dimension involves workforce transformation. AI adoption redefines job responsibilities, requiring continuous reskilling that harmonizes human expertise with intelligent systems. Workday's AI-powered recommendations and skills cloud framework are strong examples of how AI-enabled ERP platforms promote continuous learning, human–machine collaboration, and adaptive workforce management aligned with Industry 5.0 principles.

This human-centric approach fosters trust, innovation, and ethical engagement. Moreover, integrating AI with the Internet of Things (IoT) and digital twin technologies enables real-time visibility, predictive maintenance, and cross-enterprise synchronization. In sectors such as healthcare or manufacturing, where Workday connects through API-driven architectures, this integration streamlines end-to-end operations, enhancing efficiency, resilience, and resource utilization.

However, this technological advancement must be matched with stronger focus on data security, interoperability, and change management to ensure ethical scalability. Environmental and social implications also deserve greater attention, as AI and big data analytics in ERP systems create opportunities for sustainability through energy efficiency and carbon footprint monitoring. Future ERP evaluations should therefore include sustainability and social responsibility metrics alongside traditional KPIs.

Ultimately, achieving optimal integration of AI into ERP systems requires a robust governance and evaluation framework that assesses both technical accuracy and ethical alignment while preparing the workforce and business strategies for an AI-driven future. By embedding explainability, agility, and accountability into ERP design, organizations can ensure that AI-integrated systems deliver stronger audit integrity, higher user trust, and sustainable long-term value.

.

## CONCLUSION

A comprehensive review emphasizes that AI integration contributes to enhancing ERP functionalities by providing predictive improvements, automating complex tasks, and optimizing decision-making, ultimately leading to greater organizational performance and competitiveness. Nevertheless, emerging issues in data governance, ethical AI applications, and the integration of various AI models into existing ERP systems call for further investigation and the development of robust assessment frameworks. These frameworks must address not only technical effectiveness but also broader implications for organizational culture and leadership, particularly in creating a transformative environment that maximizes the impact of AI-driven ERP.

This includes developing new metrics to evaluate the adaptability and ethical transparency of AI algorithms in ERP systems, as well as providing opportunities for workforce upskilling to fully leverage these advanced technologies. The dynamic fusion of digital twins and IoT with AI within ERP architecture enables real-time monitoring, process control, and cross-enterprise synchronization in rapidly changing environments. The orchestration of these technologies holds significant potential for transforming supply chain management by minimizing disruptions and optimizing resource utilization.

At the same time, thoughtful consideration must be given to the scalability of AI solutions, along with the ethical challenges of data-driven approaches. Ensuring data security, interoperability, and effective change management is essential for the successful deployment of next-generation ERP models that promote collaboration and efficiency. The use of explainable AI tools can also enhance transparency and trustworthiness in key enterprise decisions, addressing the "black box" problem in many AI applications.

The convergence of AI, big data analytics, and IoT within the Industry 4.0 and 5.0 paradigms present new opportunities for improving sustainability, energy efficiency, and carbon footprint management in ERP-enabled manufacturing systems. This diversity, however, warrants comprehensive evaluation frameworks that incorporate environmental and social impacts alongside traditional performance indicators. Moreover, limited research exists on the interaction of different AI technologies within ERP ecosystems and their combined influence on organizational performance and resilience.

Future research should focus on developing integrated frameworks that account for the interdependencies among AI components, supporting human–machine collaboration and enhancing adaptive decision-making in dynamic business environments. Understanding how human–robot cooperation models within Industry 5.0 can be embedded in ERP systems will help optimize workflows, foster innovation, and reduce resistance to new technologies. A balanced approach that merges human expertise with AI-driven insights is necessary to achieve scalability, resilience, and ethical operation in rapidly evolving industries.

## FUTURE SCOPE

1. Development of Standardized AI-ERP Evaluation Frameworks: Future studies should focus on designing unified evaluation models that incorporate accuracy, adaptability, ethical compliance, and user trust for AI-driven ERP performance assessment.
2. Explainable AI (XAI) in ERP Decision Pipelines: Research is needed to explore how explainable AI techniques can increase interpretability and accountability in ERP decision-making, especially in finance, HR, and compliance workflows.
3. Human Centric AI Collaboration and Workforce Readiness: Further research should investigate the impact of AI-enabled ERP adoption on workforce roles, required competencies, and human centric machine collaboration models within Industry 5.0 environments.
4. Integration of IoT, Digital Twins, and AI in ERP Ecosystems: Future studies could explore how AI-connected ERP systems and interact with IoT-enabled assets and digital twins to create real-time, self-correcting enterprise environments.
5. Sustainability and Ethical Governance Metrics in AI-ERP: Further research should incorporate sustainability, environmental footprint optimization, and ethical AI governance as measurable indicators in next-generation ERP frameworks.